\documentstyle[aps,prd,epsfig,floats]{revtex} 

\draft

\def\de{decoherence }
\def\me{ L.Stodolsky}
\begin{document}
\title{%
\hbox to\hsize{\normalsize\rm 
\hfil Preprint MPI-PhT/2002-06}
\vskip 36pt Lessons of     
Coherence and Decoherence--From Neutrinos to SQUIDS}

\author{L.~Stodolsky}
\address{Max-Planck-Institut f\"ur Physik 
(Werner-Heisenberg-Institut),
F\"ohringer Ring 6, 80805 M\"unchen, Germany}

\maketitle
\bigskip
\centerline{ Presented at the XXII Solvay Conference, The Physics
of Information}
\centerline {Delphi, Nov 2001}

\begin{abstract}

 We indicate some of the lessons learned from our work on coherence
and decoherence in various fields and mention  some recent work
with solid state devices as elements of the ``quantum computer'',
 including the realization of simple logic gates controlled by
adiabatic processes. We correct a commonly held misconception
concerning  decoherence for a free particle.
\end{abstract}
\vskip2.0pc

The subject of ``quantum information'' and in particular its
realization in terms of real devices revolves in large measure
around the problems of coherence and \de. Thus it may be of
interest here to review the origins of the subject and see what has
been learned in  applications to various areas.
We first got involved in  these issues through the
attempt to see the
effects of parity violation (``weak neutral currents'') in handed
molecules~\cite{beats}. The method we found--an analogy to the
famous neutral $K$  meson behavior  with chiral molecules--
seemed too good to be true: we had a way of turning $10^{-15}$ Ev
into a big effect! There must be some difficulty, we felt. Indeed
there was; it turned out to be what we called ``quantum damping''
and what now-a-days is called ``decoherence''.

 The lessons from this work were several and interesting. First,
concerning parity violation, we realized that this could solve
Hund's ``paradox of the optical isomers'' as to why we observe
handed molecules when the true ground state should be parity even-
or- odd linear combinations. We realized that for 
molecules where tunneling between chiral isomers is small, parity
violation dominates and the stationary state of the molecule
becomes a handed or chiral state, and not a 50-50 linear
combination of chiral states. 

This holds for a perfectly  isolated molecule, and in itself has
nothing to do with \de. However, and this is very related, even a
very small interaction with the
surroundings suffices to destroy the coherence necessary for the
aforesaid linear combination, in effect the environment can
stabilize the chiral states. This now
goes under the catch-word ``\de by the environment''. The limit of
strong damping or stabilization is  often called the Zeno or
``watched pot'' effect, an idea which as far as I can tell, goes
back to Turing.  We were able to show how this just arises as the
strong damping limit of some simple ``Bloch-like'' equations
\cite{time,timea}.

\section{The Unitarity Deficit Formula}

A result of this work  is that there is a
simple and illuminating formula for the 
\de rate. There is a quantity $\Lambda$,
given by the flux of the surrounding particles or excitations,
and the $S$ matrix for the interaction of our system (e.g. the
chiral molecule) with these surroundings:

\begin{equation}\label{lam}
\Lambda= i (flux)<i\vert( 1-S_LS_R^\dagger  )\vert i>
\end{equation}

The imaginary part gives the \de rate or loss of phase coherence
per unit time $D$: 

\begin{equation}\label{d}
D= Im~\Lambda 
\end{equation}

(The real part also has a
significance, a level shift induced by the surroundings. This turns
out to be a neat way to find the index of refraction formula for a
particle in  a medium \cite{rev,nab}.)

 The labels (L,R) on the $S$ refer to  which state of the molecule
(or other system) is doing the interacting with the surroundings.
Here with (L,R) we have taken the case of the simplest non-trivial
system, the two-level system. 

These equations may be derived\cite{timea,rev} by thinking
of the S-matrix as the operator which transforms  the initial state
of an incoming object  into the final state. If
the different states (L,R) of our system scatter the object
 differently, a ``lack of overlap'' or ``unitarity
deficit'' as given by Eq~[\ref{lam}] arises. These intuitive
arguments
can also supported by more formal manipulations~\cite{nab}.

 An important point that we see here, in Eq~[\ref{lam}], is
that the environment ``chooses a direction in hilbert
space''\cite{rev}.  That is, there is some direction (here
L,R) in the internal space of the system under study (the molecule)
that is left  unchanged--is not ``flipped''-- by the interaction
with the surroundings. Such states however get a phase factor by
the interaction, and this is the \de. If the interaction did not
distinguish
some direction, if we had $S_L=S_R$ then the formula tells
us there would be no \de. This is intuitively correct in accord
with one's ideas about ``measurement''. If the probe does not
distinguish any state  there are no ``wavefunction  collapses'' and
no \de takes place. (This is not meant to imply sanctioning of 
``wavefunction collapses'' in any way.)

 Another simple limit for the formula occurs when only one state
interacts, say no interaction for L, or $S_L=1$. Then one finds
that
the \de rate is 1/2 the scattering rate for the interacting
component~\cite{timea}. Thus Eqs~[\ref{lam},\ref{d}] have two
interesting limits:

\begin{equation} \label{no} 
S_L=S_R~~~~~~~~~~~~~~~~~D=0\; ,~~no~ decoherence
\end{equation}
and
\begin{equation}
S_L=1~~~~~~~~~~~~~~D=1/2 ~(scattering~ rate~of~R)
\end{equation}

The latter followed from an application of the optical theorem.
With appropriate evaluation of the S-matrices,
Eqs~[\ref{lam},\ref{d}]  can be applied
to many types of problems, like quantum dots~\cite{vbl} or
neutrinos~\cite{thml}, or even gravity~\cite{grav}.

Eq~[\ref{no}] is quite interesting in that it says the system can
interact but nevertheless  retain its internal coherence. A lesson
here is that one shouldn't think that every interaction or
disturbance ``decoheres'' or ``reduces'' the system. The system can
interact quite a bit as long as the interactions don't distinguish
the different internal states.

 \section{A common misconception}

The fact that the interaction responsible for the \de must 
``choose a direction in hilbert space'' has some interesting
implications. One of these has to do with the \de of  a free
particle in some background environment.

 Eq~[\ref{lam}] was for a two-state system, and the extension to 
a larger number of states, as long as it is a finite number, can be
easily envisioned as following  the  logic\cite{time,timea,rev} 
used in finding
Eq~[\ref{lam}]. However if we go to the
continuum, that is if we  have a infinite number of states, the
problem becomes more subtle. The most common example of this is the
free particle which, say in the limit of an infinitely large
``box'', is described as system of
continuous, dense, levels.

 A number of  authors, in talking about this system, have
automatically assumed,  as indeed  first seems plausible,
that at long times the particle under the  influence of some
continually
interacting environment becomes totally ``decohered''; in the sense
that the density matrix of the particle $\rho(x,x')$ approaches the
situation of no off-diagonal elements,  that $\rho$ approaches a
$\delta$ function.  

Although this may seem plausible, that  under the repeated
bombardment by the surroundings the particle becomes more and more
``decohered'', it is in fact wrong

\begin{equation}
\rho(x,x')\rightarrow\sim\delta(x-x')~~~~~~~~~~~~~~~~~~~wrong
\end{equation}

Consider the simplest case, that of a thermal environment. On
general grounds we expect the particle in a thermal environment to
be described by the boltzmann factor, to be given by a density
matrix operator $\rho \sim e^{-H/T}$, where T is the temperature
and H the hamiltonian, say $p^2/2m$ for a non-relativistic
particle.
Now evaluate this operator in the position representation:

\begin{equation} \label{boltz}
\rho(x,x')=<x\vert e^{- p^2/2mT}\vert x' >\sim \int d^3p~ e^{i{\bf
p(x-
x')}- p^2/2mT} \sim e^{-(x-x')^2 MT/2}
\end{equation}

This is  the stationary, long time value of $\rho$. It
applies for nearly any state we care to initially throw into the
medium.
Evidently it shows no signs of changing and certainly no sign of
turning into a  $\delta $ function. Of course at high temperature
our expression will resemble a delta function. The practical
importance of this will depend on the other length scales in the
problem at hand. The point we wish to make, however, is of a
conceptual nature, namely that repeated interactions with the
environment don't necessarily lead to more ``decoherence''. Indeed
Eq~[\ref{boltz}] says if we  were
initially to put  $\delta(x-x') $ or some other ``highly
incoherent''
density matrix
 into the medium, the density matrix of the particle
would become {\it more coherent} with time--- until it reached the
value Eq~[\ref{boltz}]. Apparently the medium can ``give
coherence'' to a state that never had any to start with.

``Creating coherence'' by an outside influence is not as mysterious
as it may sound, there are familiar cases where we know this
already. For example, using a high resolution detector can ``create
a long wavepacket''~\cite{unn} or in particle physics neutral $K$
oscillations and the like may be enhanced or ``created'' by using
some subset of our total event sample, such as a ``flavor tag''.

Where did the seemingly plausible argument or feeling about the
indefinitely increasing \de go wrong? It's the question of the
``direction chosen in hilbert space''. The feeling is right, but we
must know where to apply it. As we can see from the
boltzmann factor, thermodynamics likes to work in momentum
(actually energy) space. The intuition would have been right
there,--in momentum space-- but this then means something
non-trivial
 in position space. 
 The lesson here is that  the notion
of ``\de by the environment'' must be understood to include a
statement about  the ``direction chosen in hilbert space'' by that
environment~\cite{rev}. 

\section{Mesoscopic systems }

 The interest in these issues has had a revival with the
advances made possible by the technologies of mesoscopic systems.
In one such system, the ``quantum dot observed by the 
QPC'', one has a complete model of the measurement process,
including the ``observer'', ``who'' in  this case is a quantum
point contact (QPC)~\cite{field}.   In a slight generalization of
the original
experiment ~\cite{buks} one can see how not only the density matrix
of the object being observed is ``reduced'' by the observing
process, but also 
see how the readout current--the ``observer'' responds. In
particular one may see how effects looking very much like the
``collapse of the wavefunction'', that is sequences of repeated or
``telegraphic'' signals indicating one or another of the two states
of the quantum dot, arise. All this without putting in any
``collapses'' by hand~\cite{vbl}.

 We should stress that what we are not only talking about  a
reduction of fringe contrast due to ``observing'' or disturbing an
interference experiment, as in ~\cite{buks}; and also in 
interesting experiments in quantum optics  where an  environment is
simulated~\cite{qopa} or different branches of the interferometer
~\cite{qopb} interact differently  and adjustably with the
radiation in a cavity (like our two S-matrices).  By the
``collapses'' however, we are referring not so much to the
interferometer itself as to the signal from some  ``observing''
system, like the current in the qpc. With repeated probing of the
{\it same} object (say electron or atom), in the limit of strong
``observation'' this signal repeats itself ----this is  the
``collapse''. For not too  strong observation there is an
intermediate character of the signal, and so on. All this may be
understood by considering the amplitude for the interference
arrangement and the readout procedure to give a certain result
~\cite{vbl}.   The properties of the readout signal  naturally 
stand in some relation to the loss of coherence or ``fringe
contrast'' of the interference effect under study.

 Following this line of thought we come to the idea that there
should be some relation between the fluctuations of a readout
signal and decoherence.
 Indeed the decoherence rate, the imaginary part of Eq~[\ref{lam}]
is a
dissipative parameter in some sense; it characterizes the rate of
loss of coherence.   Now there is the famous ``dissipation-
fluctuation theorem'',  which says that dissipative parameters are
related to fluctuations in the system. Is there some such
relationship here? Indeed, one is able to derive a relation between
the fluctuations of the readout current and the value of
$D$\cite{fltn}.
The interesting and perhaps practical lesson here is that the \de
parameter can be observed in two ways. One is the direct way, just
observe the damping out of the coherent oscillations of the system
in question. Experimentally, this involves starting the system in
a definite, selected state. However, as just explained, there is a
second way; namely  observe the fluctuations of the readout. This
can be done even if the system is in the totally ``decohered''
$\rho\sim I$ state.

Another mesoscopic system, the SQUID and in particular the rf
SQUID, has been long discussed\cite{leggett} as a candidate 
for showing that even macroscopic objects are subject to the rules
of quantum mechanics.  The rf   SQUID, a Josephson device where a
supercurrent goes around a ring, can have two distinct  states,
right- or left- circulation of the current. These two conditions
apparently
differ greatly, since a macroscopic number of
electrons change  direction. It would be a powerful argument for
the universality of
the quantum rules if one could demonstrate the meaningfulness of
quantum linear combinations of these two states.

Such linear combinations can  in principle be produced since
there is some
amplitude for a tunneling between the two configurations. In fact
this was
recently manifested  through the observation of the ``repulsion of
levels'' to be anticipated if the configurations of opposite
current do behave as quantum states\cite{luk}.

 Another approach, where we would
directly ``see'' the meaningfulness of the relative quantum phase
of the two configurations, is  the method  
of ``adiabatic inversion''~\cite{squid,squida}. This
 method also offers the possibility of a direct
measurement of the \de time.
 In adiabatic inversion the ``spin'' representing a two-level
system~\cite{timea},~\cite{thml} is made to ``follow'' a slowly
moving ``magnetic field'' (meant  symbolically, as an analogy to
spin precession physics), which is swept from ``up'' to ``down''.
 In this way the system can be made to invert its direction in
``spin
space'', that is to reverse states and go from one direction of
circulation of the current to the other. This inversion is an
intrinsically
quantum phenomenon. If it occurs it shows that the phases between
the two configurations were physically meaningful and that they
behave quantum mechanically . This may be dramatically manifested
if we  let  \de destroy the phase relation between the two
configurations. Now the configurations act classically and  the 
 inversion is blocked. 

 We thus  predict that when the \de rate is low the
inversion takes place, and when it is high it does not. 
Figs 1 and 2 show the idea of this procedure.

\begin{figure}[h]
\epsfig{file=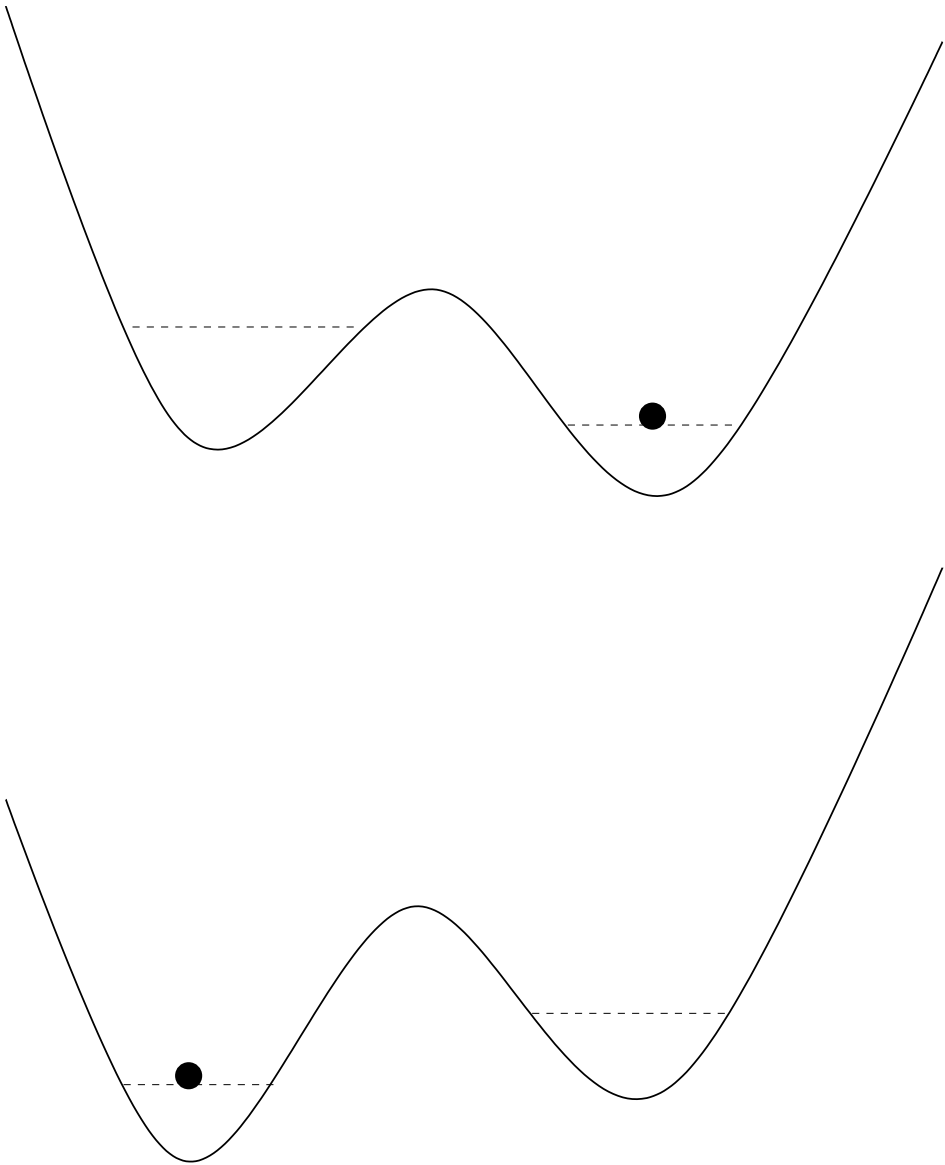, width=.7\hsize}
\hskip.3cm
\caption{ A successful inversion, starting from the upper figure
and
ending with the lower figure. The black dot indicates which state
is occupied. The system starts in the lowest energy level and by
staying
there, reverses states. It behaves as a quantum system
with definite phase relations between the two configurations. }
\end{figure}

\begin{figure}[h]
\epsfig{file=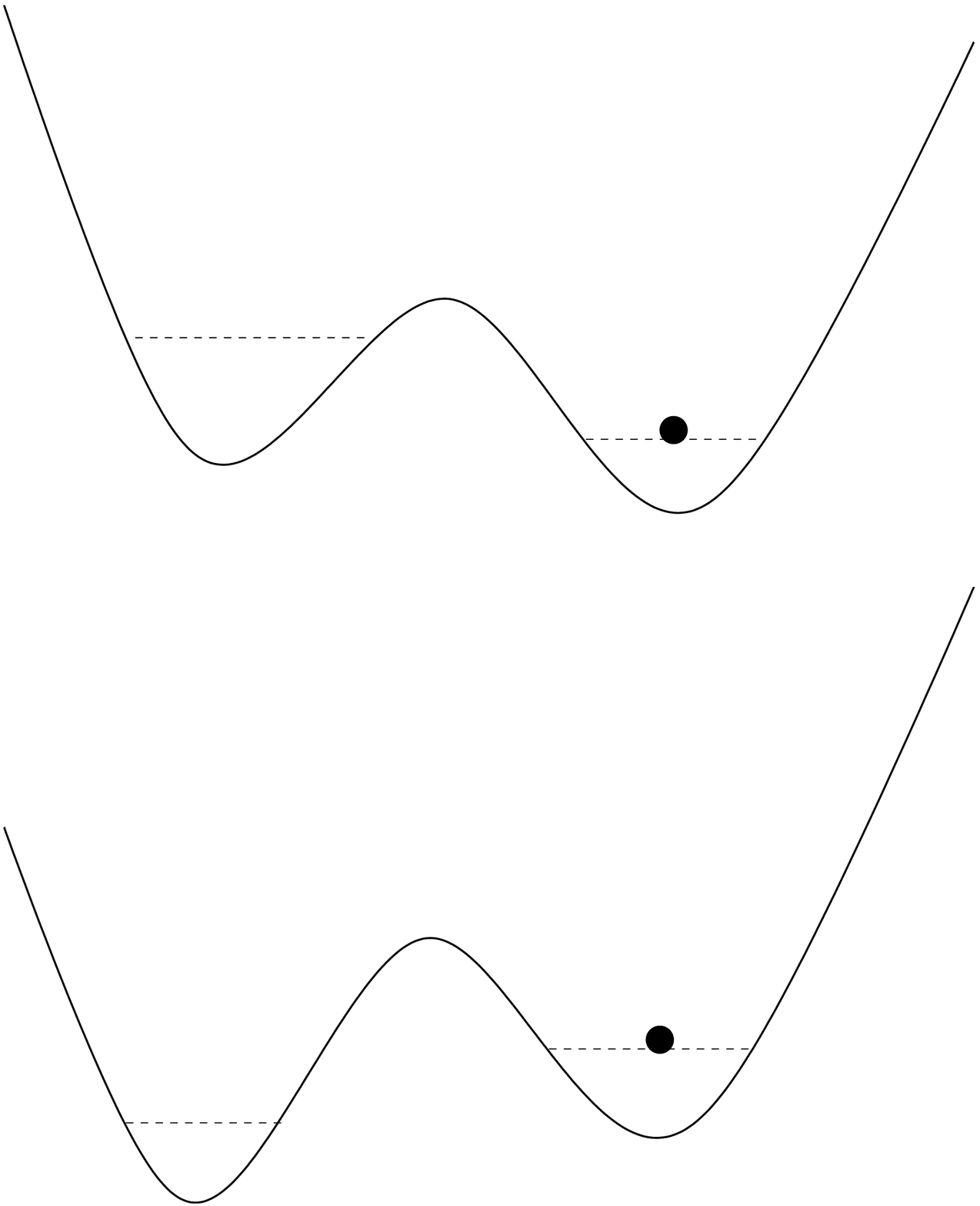, width=.7\hsize}
\caption{ An inhibited inversion, starting from the upper figure
and ending with the lower figure. Due to the lack of phase
coherence the system
 behaves classically and  stays in the same state, the current  is
not
reversed.}
 \end{figure}

 Since  in such an experiment  we  have the sweep speed
at our disposal, we have a way of determining the \de time. It is
simply the slowest sweep time for which the inversion is
successful.
We must only be sure that for the sweep speeds  in question the
the conditions remain adiabatic.

Setting up the adiabatic condition and taking some estimates for
the \de time, it appears that the various requirements can be met
~\cite{squid},~\cite{squida} when operating at low temperature.
Hence
it may be 
  realistically possible to move between the classical and quantum
mechanical worlds--to turn quantum mechanics ``on and off'' in one
experiment. This would be a beautiful experiment, the main open
question being if the estimates of the \de rate are in fact
realistic, since we are entering a realm which has not been
explored before.

\section{Adiabatic logic elements and the  quantum computer}

A two-state system behaving quantum mechanically can serve as the
physical embodiment of a quantum mechanical bit, the ``qbit''.
Furthermore, the adiabatic inversion procedure just described
amounts to a quantum realization of one of the basic elements of
computer logic: the NOT. If one configuration is identified as 1
and the other as 0, then the inversion turns a linear combination
of 1 and 0 into a linear combination of 0 and 1 with reversed
weights. 

We can try to push this idea of ``adiabatic logic'' a step further.
  NOT was a one bit operation. The next most complicated logic
operation is a two bit operation, which we may take to be 
``controlled not'' or  CNOT. In CNOT the two bits are called the
control bit and the target bit, and the operation consists of
performing or not performing a NOT on the target bit, according to
the state of the control bit. 

  To realize CNOT, an idea which suggests itself~\cite{squid} as a
generalization of adiabatic inversion is
the following. We  have a  two bit operation and so two SQUIDS.
These are devices with magnetic fields.  Now if one SQUID, the
target bit,  is undergoing a NOT operation, it can be influenced by
the control bit,  a second nearby SQUID,  through its linking flux.
We could  imagine that this linking flux can be arranged so that it
helps or hinders the NOT operation according to the state of the
second SQUID. This would amount to a realization of ``controlled
not'', again by means of  an adiabatic sweep.

 To analyze this proposal we must set up the two-variable
Schroedinger equation describing the two devices and their
interaction. The result is a hamiltonian with the usual kinetic
energy terms and a potential energy term in the two variables,
which in this case are the fluxes in the SQUIDS, $\phi_1,\phi_2$:
 
\begin{equation}
\label{pot}
V={1\over 2}V_0\bigl\{[l_1(\phi_1-\phi^{ext}_1)^2
+l_2(\phi_2-\phi^{ext}_2)^2-
2l_{12}(\phi_2-\phi^{ext}_2)(\phi_1-
\phi^{ext}_1)]+\beta_1f(\phi_1)+\beta_2f(\phi_2) \bigr\} \; .
\end{equation}
The $\phi^{ext}$ are external biases which in general will
be
time varying. The $l's$ are dimensionless inductances and $l_{12}$
represents the coupling between  the two devices. The $f(\phi)$ are
symmetric functions starting at one and decreasing with increasing
$\phi $ so as to produce a double well potential when combined with
the quadratic term; in the SQUID  $f(\phi)=cos(\phi)$.
Fig. 3 shows this ``potential landscape'' for some typical values
of
the parameters.

\begin{figure}[h]
\epsfig{bbllx=100bp,bblly=140bp,bburx=400bp,bbury=430bp,    
 file=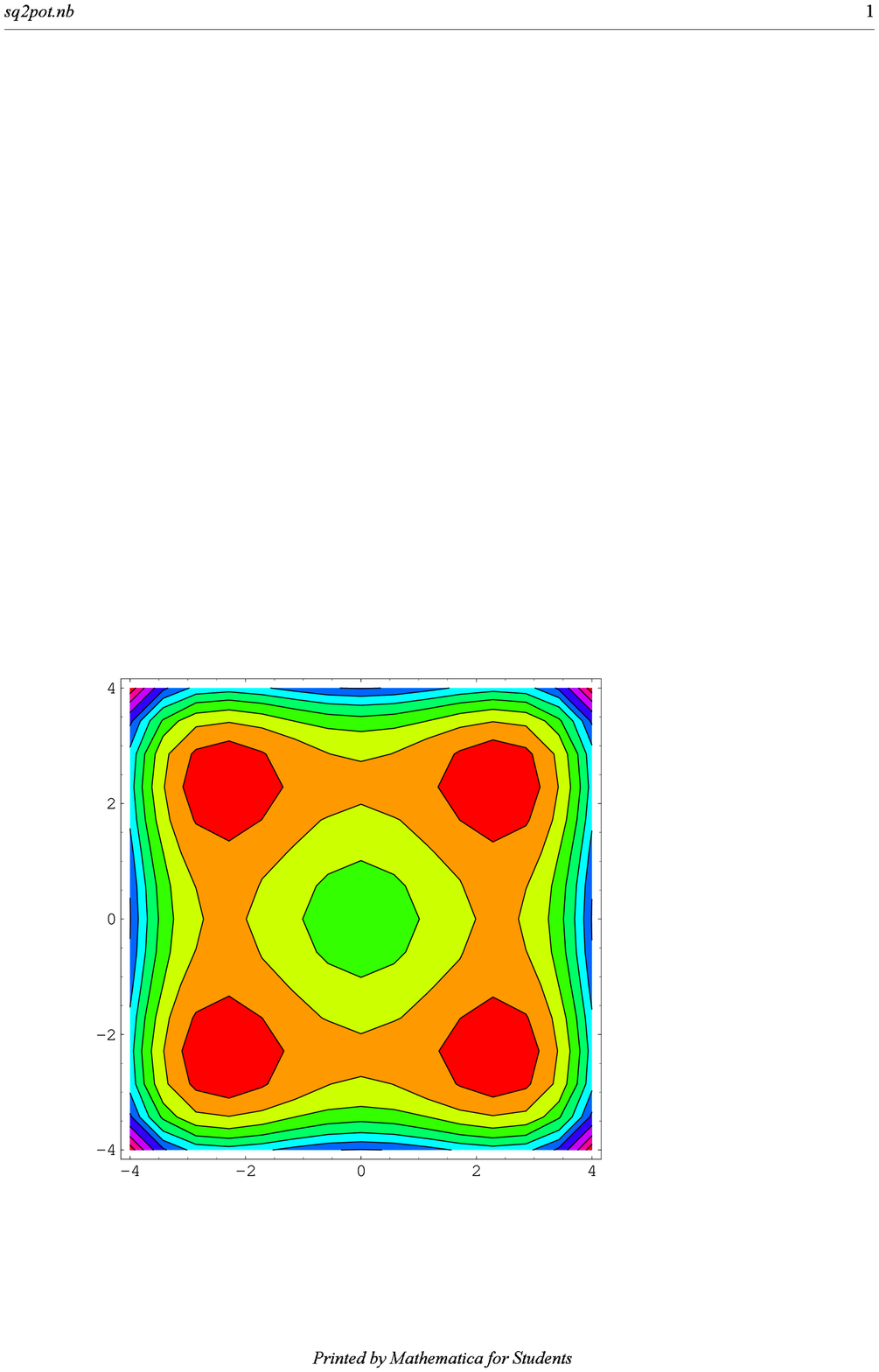, width=\hsize,clip=}
\caption{ Potential landscape for CNOT }   
\end{figure}

Given the hamiltonian, we must search for values of the
control parameters $\phi^{ext}$, the ``external fields'', which can
be adiabatically varied in such a way as to produce CNOT.   
Preliminary analysis   indicates  favorable regimes of
the rather complex parameter space where this can in fact be
done~\cite{cnot}.

\section{Some Experimental Proposals}

 Finally we would like to recall that there are still some
fundamental and beautiful experiments waiting to be done in these
areas.

 A) One is the demonstration of the large effects of parity
violation for appropriately chosen and contained handed
molecules~\cite{beats}. Because of what we now call \de this seemed
very remote at the time. But now with the existence  of single
atom/molecule traps and related techniques, perhaps it's not so
hopeless.

B) Another, concerned with fundamentals of quantum mechanics, could
be called the ``adjustable collapse of the wavefunction'' where the
``strength of observing'' can be varied, leading to effects like
washing out of interferences, as already seen in \cite{buks}  and 
a number of further predictions where we vary the qualities of the
``observer'' \cite{vbl}, or slowing down of relaxation according to
the rate of probing of the object~\cite{timea}. 
 
C)  Then there is the second way of measuring $D$, through the
fluctuations in the readout signal, even for a fully ``decohered''
system. 

 D) Finally there is the direct demonstration of quantum linear
combinations of big objects by the method of adiabatic inversion; 
 ``turning quantum mechanics  on and off''~\cite{squid,squida}. 

Many of the questions we have briefly touched upon had their
origins in an unease with certain consequences of quantum
mechanics, often as ``paradoxes'' and ``puzzles''. It is amusing to
see how, as we get used to them, the ``paradoxes''  fade and yield
to a more concrete understanding, sometimes even with consequences
for practical physics or  engineering. If we  avoid  overselling
and some tendency to an inflation of vocabulary, we can anticipate
a bright and interesting future for ``applied fundamentals of
quantum mechanics''.


\begin{references}


\bibitem{beats} Quantum Beats in Optical Activity and Weak
Interactions,
    R.A. Harris and \me, Phys Let B78 (1978) 313.
   [Molecular tunneling and parity violation,
need to understand decoherence.]

\bibitem{time} On the Time Dependence of Optical Activity,
    R.A. Harris and \me, 
   J. Chem. Phys. {\bf 74} (4), 2145 (1981).[Decoherence by
environment, formula for decoherence rate, application to chiral
molecules.] The notion that handed molecules could ``decohere'' or 
be stabilized by the environment somehow was raised by H.D.Zeh,
Found.
Phys. 1,69 (1970), M. Simonius, Phys Rev. Lett. 40, 980 (1978).

\bibitem{timea} Two Level Systems in Media and `Turing's Paradox',
     R.A. Harris and \me, Phys. Let. B 116(1982)464.[Decoherence by
environment, formula for decoherence rate, quantitative explanation
of ``Zeno'', prediction of anti-intuitive relaxation, application
to neutrinos.]

\bibitem{thml} On the Treatment of Neutrino Oscillations in a
   Thermal Environment, \me,  Phys. Rev. D 36(1987)2273 .[Method
for
decoherence
in neutrino oscillations. Spin precession  picture for neutrinos.]
See chapter 9 of G.G.~Raffelt,
{\it Stars as Laboratories for Fundamental Physics} 
(Univ. Chicago Press, 1996)

\bibitem{unn} When the Wavepacket is Unnecessary, \me, Phys. Rev.
{\bf D58}  036006, 1998 and  hep-ph/9802387.

\bibitem{rev} Quantum Damping and Its Paradoxes,\me,
in {\it Quantum Coherence}, J. S. Anandan ed.
World Scientific, Singapore (1990)[Review and historical background
to 1989.]



 
 \bibitem{nab} Non-Abelian Boltzmann Equation for Mixing and
Decoherence
 G. Raffelt, G. Sigl and \me, Phys Rev Let.
70(1993)2363.[Derivation of
decoherence rate formula in neutrino applications ]

\bibitem{grav} Decoherence Rate of Mass Superpositions, \me,
   Acta Physica Polonica {\bf B27},1915(1996). [Decoherence in
gravitational interactions.] 


 
\bibitem{buks}
S. Gurvitz, quant-ph 9607029.
 E. Buks, R. Schuster, M. Heiblum, D. Mahalum and V.
Umanksy, Nature {\bf391}, 871 (1998).

\bibitem{field}
M. Field, C.G. Smith, M. Pepper, D. A. Ritchie, J. E. F. Frost G.A.
C. Jones and D.G. Hasko,  Phys. Rev. Lett. {\bf 70}, 1311 (1993). 




\bibitem{vbl} Measurement Process In a Variable-Barrier System,\me,
Phys. Lett. {\bf B459} pages 193-200, (1999). [Formalism for
quantum dot-QPC system. Prediction of novel phase effect.
Prediction of ``collapse-like'' behavior of readout.]

\bibitem{qopa} Myatt et al. Nature 403, 269 (2000).

\bibitem{qopb} Brune et al. Phys. Rev. Lett. 77, 4887 (1996).


\bibitem{fltn} Decoherence-Fluctuation Relation and Measurement
Noise,\me,
Physics Reports {\bf 320} 51-58 (1999), 
quant-ph/9903075. [Suggestion of Decoherence-Fluctuation Relation
connecting decoherence rate and fluctuations of readout signal.] 

\bibitem{leggett} See A. J. Leggett, Les Houches, session XLVI
(1986) {\it Le Hasard et
 La Matiere}; North -Holland (1987),  references cited therein, and
introductory talk, Conference on Macroscopic
Quantum Coherence and Computing, Naples, June 2000,
Proceedings published by Academic-Plenum.)

\bibitem{luk}
 J. Friedman, V. Patel, W.Chen, S.K. Tolpygo and J.E. Lukens,
  Nature {\bf 406}, 43 (2000).  C.S. van der Wal, A.C. ter Haar, F.
K. Wilhelm, R. N.
Schouten, C.J.P.M. Harmans, T.P. Orlando, Seth Lloyd, and J.E.
Mooij, Science {\bf 290} 773 (2000) and
  in MQC2: Conference on Macroscopic Quantum Coherence and
Computing,
Naples, June 2000,
Proceedings published by Academic-Plenum.
 
\bibitem{squid}
Study of Macroscopic Coherence and Decoherence in the SQUID by
Adiabatic Inversion, Paolo Silvestrini and \me,
Physics Letters {\bf A280} 17-22  (2001).[Linear Combinations of
macroscopic states. Measuring decoherence time. Relation to NOT
operation.] cond-mat/0004472 

\bibitem{squida}
Adiabatic Inversion in the SQUID, Macroscopic Coherence and
Decoherence,   Paolo Silvestrini and \me,  {\it Macroscopic Quantum
Coherence and Quantum
Computing}, pg.271,  Eds. D. Averin, B. Ruggiero and P.
Silvestrini, Kluwer Academic/Plenum, New York (2001). [Linear
Combinations
of macroscopic states. Measuring decoherence
time.]cond-mat/0010129. D.V. Averin, Solid State Communications
{\bf 105},
659 (1998),
 has discussed related ideas using adiabatic operations on
the charge states of small Josephson junctions.

 
\bibitem{cnot}  Adiabatic Methods for a Quantum CNOT Gate,
Valentina Corato, Paolo Silvestrini,\me, and Jacek Wosiek,
 cond-mat/0205514, and contribution to {\it Macroscopic Quantum
Coherence and Quantum
Computing 2002}. [ Principles of quantum gates using adiabatic
inversion. Design parameters for CNOT.] 
\end{references}
\end{document}